\documentclass[12pt]{article}
\usepackage{graphicx}
\usepackage{amsfonts}
\usepackage{amscd}
\usepackage{latexsym}
\usepackage{epsf}

\newcommand{\be}{\begin{equation}}
\newcommand{\ee}{\end{equation}}
\newcommand{\bea}{\begin{eqnarray}}
\newcommand{\eea}{\end{eqnarray}}

\newcommand{\nn}{\nonumber}
\newcommand{\fr}{\frac}
\newcommand{\pd}{\partial}
\newcommand{\al}{ \alpha }

\newcommand{\ep}{ \epsilon }

\newcommand{\ra}{\rightarrow}

\usepackage[latin9]{inputenc}
\usepackage{textcomp}
\usepackage{amsmath}
\usepackage{multirow}
\usepackage{color}
 
\newcommand {\si} {\sigma}

\newcommand {\ga} {\gamma} 
  
\newcommand {\pb} {\bar{p}}  
\newcommand {\rb} {\bar{r}}  
\begin{document}
\title{\bf A Model of Hydrogen Desorption Kinetics Controlled both by Interface and Surface Reactions 
 for Metal Hydrides}
\author{I.  Drozdov\\}
\date{ }
\maketitle

\begin{abstract}
 The desorption kinetics was modelled with the both interface- and surface
 reactions as rate-controlling steps. 
 It has been shown analytically, that in the model of 'shrinking core' desorption,
 the finite hydride-decomposition-rate causes a modified slope of kinetics.
  The dependence of desorption time on the powder particle size has the same 
 power of order as for the surface controlled desorption.
\end{abstract}


\section{Introduction}

Metal hydrides considered as a possible hydrogen storage,  
possess a very high hydrogen capacity.
 A common disadvantage of metal hydrides for practical applications is  a 
relatively unsufficient  desorption kinetics.

The sorption mechanism is outlined below for the example of magnesium hydride $MgH_2$.
 The uptake of high amount of hydrogen becomes possible through
 the formation of a hydride stoichiometric structure ($\beta$-phase) in the metal with
 single dissolved hydrogen atoms ($\al$-phase).

 Several processes governing the kinetics of absorption and
desorption occur in the following stages:

 {\bf a)} surface ($\al$-surface) adsorption/desorption - the chain of reactions on the outer surface providing the
transition between the molecular hydrogen gas and dissolved hydrogen atoms in the metallic lattice.

{\bf b)} interface ($\beta$-surface) formation/decomposition - the transition between the $\al$-dissolved hydrogen atoms 
in metallic magnesium and the stoichiometric hydride on the surface of $MgH_2$-grain.

{\bf c)} a transport of hydrogen atoms  from the one to another surface by thermodynamical diffusion
 (only through the concentration gradient without other driving forces).

 Obviously, processes of these three stages cannot influence the sorption kinetics independently
from each other, since the rate of each next one is strongly coupled with the rate of 
the previous one. The resulting kinetics is determined therefore by the slowest process of the 
entire chain and is called the 'rate controlling step'.
 The question of the rate controlling step is crucial to understand the mechanism of kinetics for   
 systematical improvement of sorption properties.
 
 An apparently simple relation seems to exist between the sorption kinetics and the particle 
size, related to the specific surface of powder. It is proved experimentally, that a refinement of powder
particles through e.g. mechanical milling, increases the sorption (especially desorption) rate \cite{}.

 Several models for ad-/desorption based on the `shrinking core' scheme, were recently investigated, 
assuming processes on the surface ({\bf a)}) only \cite{1stpaper, 2stpaper}; on surface +
bulk diffusion ({\bf a)+ c)}) \cite{diffusion_paper}
 In the present model we suppose the rate controlling step to be the interface  process (hydride decomposition),
 which occurs nearly as fast as the process on the surface ({\bf a)}+ {\bf b)}).
The stage {\bf a)} can be never withdrawn from the consideration, since the
 interchange between surrounding gas and metal surface is strongly subjected to the Sievert's
law, which is proved by existence of a threshold pressure both for ad- and desorption.
 Nevertheless, the diffusion {\bf c)} is assumed to be significantly faster, than both {\bf a)} and {\bf b)}. 
 
The sorption kinetics controlled by each of these different stages should
 have also different characteristic powers from pure dimensional reasons. So, the desorption
 time $\tau$ (1/rate) {\it controlled by the surface } should be {\it linear} in the particle size $L$    
  at isobaric desorption \cite{1stpaper}. 
The finite reaction rate of the hydride decomposition, which is the surface reaction running
on the interface ( surface of $\beta$-core) does not change this linear tendency. The delaying 
influence of this process results entirely in the modified slope of the desorption kinetics, which
is also different for middle and final phase of the desorption.
  
 The degree of this influence can be estimated by a simplified analytical modelling, as it
is performed in the next sections.  

In contrast, the {\it diffusion controlled} kinetics  increases quadratically in the $L$, as it was shown
 recently in \cite{diffusion_paper}.
 
\section{Interface Reaction of a Finite Rate}

\subsection{ Theoretical foundations of the model }

A number of models \cite{...} based on a pure phenomenological preset, attempt to get the
 desorption/absorption kinetic rate as the dependence of the sorption amount, as it should be then proved 
experimentally.
 On the other hand it is clear, that every chemical/physical process which occurs for example 
on the surface or  interface, is a strictly {\it local} process
and its kinetics in a certain point (infinitesimally small volume) can be only dependent
 on values in the same point. The rate of $MgH_2$-decay ($\beta\ra\al$) transition in the
point of the interface and the velocity of this surface resulting thereby, can be only dependent
on temperature, elastic stress, concentration of $\al$-dissolved hydrogen, gradient of the concentration,
 diffusion flow, eventually also higher derivatives of concentration, geometrical properties
 of the surface (curvature, lattice orientation etc.) taken in this point. 
 The integral representation of this local behavior can lead to the resulting dependence
on {\it global} measured parameters, such as the desorbed hydrogen amount or the ratio
hydrogen/metal in a single powder particle or in the sample. 

 In this sense, we suppose the $\mu$ to be the mass/molar rate of the decomposed magnesium hydride 
on the interface - '$\beta$-surface decomposition rate'- that means the mass/molar amount of magnesium hydride 
decomposing during the time $dt$ per interface area $d\si$:
\be 
\mu:=\fr{dm_{MgH_2}}{dt\ d\si}=\mu(c_\al, \nabla c_\al, ...  );\  \left[ \fr{kg}{m^2\cdot s} \right],
\label{mu_definition}
\ee
as assumed. The rate of hydrogen released in the surrounding $\al$-solution can reach theoretically
$\ga\mu$ maximal, $\ga:=0.0766$. 

Since the variable concentration of hydrogen atoms is only the concentration of the $\al$-dissolved
 hydrogen $c_\al$, we omit further the subscript $\al$ of it, $c \equiv c_\al$.
  The function $c$ is now considered as a dynamic scalar field $c=c(\vec{r},t)$ in the subdomain
of non-stoichiometric $\al$-phase, i.e. in the space between the interface from inside and surface from outside.

 It should be pointed out, that a simple diffusion law in the Fick's form 
\be 
{\bf j}=-D\nabla c, 
\label{fick}
\ee
does not hold in the vicinity of the surfaces because of local 
surface effects. We define therefore this area for the $\beta$-surface as some layer which belongs to it.
 Thus, the interface as a boundary in the $\al$-domain is understand as a boundary surface, up from which the law \ref{fick} is valid.  

The molar rate of dissolved hydrogen transported by diffusion away from this interface through 
the $\al$-domain per unit surface occurs according to the (\ref{fick}), 
where the coefficient $D$ in every point may be generally dependent on concentration and other local variables.

 As assumed above in the introduction, in the present approach we consider the diffusion rate
to be faster than surface reaction rates {\bf a} and {\bf b}. Then the concentration profile
can be considered to be `quasi-stationary' \cite{diffusion_paper}.
In the approach of {\it quasi-stationary} concentration profile
we assume the concentration to obey in each point of $\al$-domain in any
 time the stationary diffusion equation (Laplace equation). 
  The time-dependence of $c$ comes about from the time-dependent boundary condition on the interface.         
  
 Further simplification results from the radial symmetry of the model. 
 We relate the center of radial coordinate system to the central point of radial symmetric
particle (ball) of a constant radius $L$ with the radial symmetric $\beta$-core of a variable radius $\rho$,
 like it has been made in a number of similar models \cite{gabis, castro, 1stpaper, diffusion_paper}.
 The unique nontrivial solution of the radial Laplace equation 
\be \fr{\pd^2}{\pd r^2}c+\fr{2}{r} \fr{\pd}{\pd r}c = 0  \ee
is the ansatz
\be  c(r) = \fr{A}{r}+B  
 \label{ansatz}
\ee
with some constants $A,B$ whereat the case $A > 0$ corresponds to desorption, $A < 0 $ to absorption,
 respectively.

The ansatz (\ref{ansatz})is fixed at $\al$ surface by the boundary condition:
\be -\left.D_\al \nabla c \right|_{r=L}= D A/L^2 =b c(L)^2 - k p, \ee
which is the Sievert's law, modified by the surface re-adsorption factor $k$,
as introduced by \cite{gabis} and explained detailed in \cite{1stpaper}. For the radial coordinate
$r$ it reads
\be
  D_\al c'(r)= bc(r)^2 -k p\  |_{r=L}
\label{radial_boundary}
\ee

\subsection{A Constant Interface Reaction Rate}

 The evolution of the interface is governed therefore by the diffusion flux and the surface reaction,
represented by the function $\mu$.

 We assume for a first approximation, the chemical reaction on the interface 
$ MgH_2 \ra Mg + 2H$ occurs with the constant rate $\mu $ dimensionalized e.g. as 
$[mol/(m^2\cdot s)]$. 
 The total rate of the atomic hydrogen released into the surrounding metallic magnesium is
proportional to the total interface area $\si,\ [m^2]$.
 To consider the rate $\mu$ as a possible rate controlling step, we should suppose the bulk 
diffusion rate $D$ to be anyway much faster than $\mu$.   

 For a case of {\it radial} symmetry we have then for the quasi-stationary concentration $c$ of atomic 
hydrogen  $ c(r)=A/r+B $ on the inner surface:
\be
D\fr{A}{\rho^2}=\mu,
\ee
 and for the hydrogen balance 
\be
 [Y-c(\rho)] \fr{d \rho}{d t}=\left[ Y-\left( \fr{A}{\rho}+ B \right) \right] \fr{d \rho}{d t}=\mu
\ee
where $A,B$ are now functions of $\rho$.
For the outer surface $r=L$ we have still
\be
\fr{A}{L^2}= b\left( \fr{A}{L}+B \right)^2-k p,
\ee  
according to \ref{radial_boundary} with the constant external pressure $p$.
These three equations together provide the evolution equation for the relative
$\beta$-core radius $\rb=\rho/L$:
\be
\left(  Y-\al \sqrt{\rb^2+\beta }\right) \fr{d\rb }{d t} = \fr{\mu}{L} 
\ee 
  $ \al=\sqrt{\mu/b};\ \ \ \beta= k p/\mu  $,
with an analytical solution
\be
\fr{\mu}{L}t= Y(\rb-1) -\sqrt{\mu/b} \left[ \rb\sqrt{\rb^2 +\beta } -\sqrt{1 +\beta } 
+ \beta \ln \fr{\rb+\sqrt{\rb^2+\beta} }{ 1+\sqrt{1+\beta} }  \right].
\ee

 The total desorption time $\tau$ corresponding to the $\rb = 1$ is
\be
\tau= \fr{L}{\mu} \left[-Y + 
\sqrt{\fr{\mu}{b}}\left(  \sqrt{1+\beta}+\ln \fr{ 1+\sqrt{1+\beta} }{\sqrt{\beta}}  \right)\right]
\ee

\subsection{The Linear Concentration-Dependence of Inner Surface Reaction Rate}

The hydride decomposition with a constant rate  $\mu$ proposed above, allows to consider this
reaction as a rate controlling step and can also describe appropriately the experimental data
for small values of $\mu$. In general, for an arbitrary $\mu$, the assumption $\mu=$const is not consistent
 with the physical reality anymore.
  The reason is the existence of the highest possible concentration $c_{max}=X $ (for a given temperature)
 of the atomic hydrogen in magnesium ($\al$-dissolution). 

 As considered previously in \cite{1stpaper, 2stpaper, diffusion_paper}, the molar concentration $c$ of
 $\al$-dissolved hydrogen atoms can never overcome 
the critical value $X [mol/m^2]$. 

In the simplified case, the rate $\mu$ is the function only of the $\al$-concentration.
Once the critical concentration $X$ is reached, the 
hydride decomposition stops (saturation). 
The decomposition rate $\mu$ is dependent on the concentration, so that $\mu(c=X)=0$. Different forms
of this function are possible (Fig.1 a), b) ). This shape is temperature-dependent and generated entirely by the general
structure of the chemical potential.\\

\includegraphics[scale=0.33]{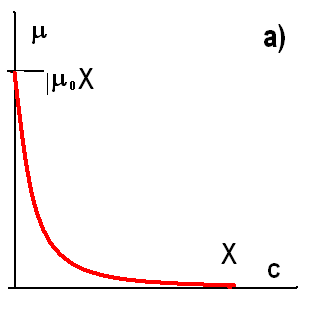}\hspace*{0.5cm}
\includegraphics[scale=0.33]{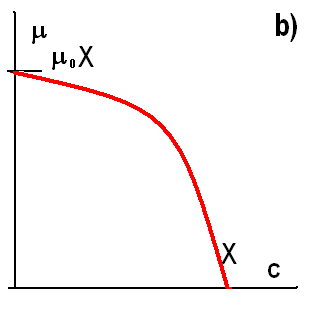}\hspace*{0.5cm}
\includegraphics[scale=0.33]{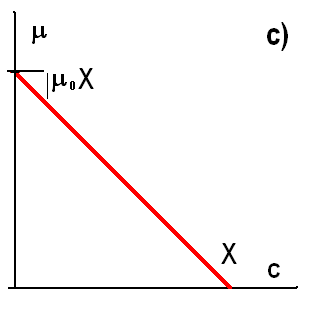} {\bf\small Fig.1}\\

 The simplest model 
reproducing this behaviour is the linear descent from $\mu_{max}=\mu_0 X$ at $c=0$ to $\mu=0$ at $c=X$ 
of the form: $\mu(c)=\mu_0 (X-c)$, with $0 < c < X$ (Fig.1 c)).
 The constant $\mu_0$ has now a dimensionality of velocity, that should not be confused with the physical boundary
velocity.   
The balance and boundary conditions on $ r=\rho, r=L $ take the form. 
 \bea 
\small
\left[ Y -\left( \fr{A}{\rho} + B \right) \right]\fr{d\rho}{d t} & = &
 \mu_0  \left[ X- \left( \fr{A}{\rho} + B \right)\right]\\
\fr{D A}{\rho^2} & = & \mu_0 \left[ X- \left( \fr{A}{\rho} + B \right)\right]\\
\fr{D A}{L^2} &= & b\left( \fr{A}{L}+B \right)^2-k p
\eea
 Further, with the notations
\be
y:=X-\left( \fr{A}{\rho}+B \right);\  \  \rb:=\rho/L; 
\ee
we obtain the solution in the form:
\be
 \mu_0 t = -L\rb |_1^{\rb} + L(Y-X) \int\limits_1^{\rb}  \fr{d\rb}{y},
\label{integral}
\ee
where $y$ is obtained from boundary conditions as the function:
\be
\small
y=\fr{ X-\fr{\mu_0}{2b}\left[ \sqrt{\ga^2+\fr{4b}{\mu_0^2}\left( k p +\mu_0\ga X\right)} - \ga \right]}{\rb^2/\ga}
\mbox{ with }
\ga =\ep \fr{\rb^2}{\rb(1-\rb)+\ep}, \ \ \ep=\fr{D}{\mu_0 L}
\label{y_ga}
\ee

 We remind on the assumption of a fast diffusion $D$, compared to the reaction rate $\mu_0$, and furthermore
we take into account the size of powder particles $L\sim 10^{-6}..10^{-8}$ m. Thus we have for the dimensionless
relative rate $\ep >> 1$, whereas the $\rb$ for the main desorption phase is typically of order unity. 
 This fact allows for replacement 
\be
\ga\ra \rb^2\  \mbox{  by the expansion }\  \ga\approx \rb^2-\fr{\rb^3}{\ep} + O(\rb^4). 
\ee
 It reduces the solution (\ref{integral}-\ref{y_ga}) to the form:
\bea
\mu_0 t &=& L(1-\rb)+(Y-X)L\fr{2 b}{\mu_0} \int\limits_1^{\rb} \fr{d\xi}{a+\xi^2 - \sqrt{\xi^4+2 a\xi^2+c }}\nn\\
\mbox{ with } a:&=& \fr{2b X}{\mu_0},\ \ c:= \fr{4 b k p}{\mu_0^2},
\label{integral_reduced}
\eea
and $\rb$ is renamed by $\xi$ as an integration variable.
  
 Now, a suitable comparison of reaction rates $ b, k, \mu_0$ should be considered in order to construct
an appropriable simplification of the integral (\ref{integral_reduced}). Otherwise a consequent analytical
 integration results in a very cumbersome form.    
Keeping in mind, $\xi$ remains to be of order unity, there are two possibilities:\\

i) $ \xi^2 >> a$ and, as a corollary $ \xi^4 >> c $, it means, the outer pressure is very low and
constant (the "main" desorption regime). Then 
\be
\sqrt{\xi^4 +(2a\xi^2 + c)} \approx \xi^2 + a +\fr{c}{2\xi^2}. 
\ee 
 Applied to the (\ref{integral_reduced}) it provides the solution:
\be
t=L\left[  \fr{1-\rb}{\mu_0} +(Y-X)\fr{1-\rb^3}{3 k p} \right]\\
\label{solution_i}
\ee

ii)$ \left( \xi^2 +\fr{2 b X}{\mu_0} \right)^2=(\xi^2+a)^2\ >>\ c-a^2  = \fr{4 b}{\mu_0^2}k (p-\pb) $
where $\pb:=bX^2/k$- the Sievert's threshold pressure, as considered in \cite{1stpaper}.
 This is so-called "subthreshold" regime. It provides
\be
\sqrt{(\xi^2+a)^2+c-a^2} \approx \xi^2 + a + \fr{c-a^2}{2(\xi^2+a)}.
\ee

Substituted in the (\ref{integral_reduced}), it leads to the final result:
\be
t=L\left[  \fr{1-\rb}{\mu_0}\left( 1-2 \fr{Y/X-1}{p/\pb-1}\right) +\fr{1-\rb^3}{3 k }\cdot \fr{Y-X}{p-\pb} \right]\\
\label{solution_ii}
\ee
 
\section{Conclusion}   

 The assumed model of hydrogen desorption from magnesium hydride is based on the shrinking core
 scenario. 

 Additionally to the recent investigation \cite{1stpaper}, the finite hydride decomposition rate on the
 interface has been considered as a possible rate controlling step,
 it means, this reaction is assumed to run comparable to the reaction on the outer surface.    

 The resulting kinetics of the $\beta$-core shrinkage is established analytically.
Two cases has been considered - the constant reaction rate, and the linearly concentration-dependent one,
which appears to be rather physically relevant. The latter model assumes an existence of a maximal 
concentration (saturation), as several ones considered before.
 
 It has been recovered, the slope of the desorption kinetics governed by the degassing reaction
on the surface, should be modified by the decomposition reaction on the interface,
once the latter is comparably slow. 

  The increasing of the {\it desorption} kinetic rate by refinement of particles should be referred to
 decreasing particle size, according to suggestions of \cite{1stpaper, 2stpaper}.  
 The dependence of the desorption time on the particle size remains linear.

\end{document}